\documentstyle[12pt]{article}
\newcommand{\be}{\begin{equation}}
\newcommand{\ee}{\end{equation}}                                               
\newcommand{\bear}{\begin{eqnarray}}
\newcommand{\ear}{\end{eqnarray}}
\newcommand{\lb}{\Lambda}
\newcommand{\om}{\omega}
\newcommand{\omm}{\om _{i}^{\,ab}}
\newcommand{\pve}{A_{\mu}^{\,\,ab}}

\newcommand{\psb}{\bar{\psi}}
\newcommand{\lag}{{\cal L}}
\newcommand{\act}{{\cal S}}
\newcommand{\pt}{\partial}
\newcommand{\partiall}{\partial\hspace{-6pt}/}
\newcommand{\DI}{D\hspace{-8pt}/}
\newcommand{\inte}{\int d^{3}x}
\newcommand{\dint}{d^{3}kd^{3}l}
\newcommand{\inta}{\int d^{3}kd^{3}l}
\newcommand{\vp}{\varphi}
\newcommand{\doub}{\pt _{\mu}\vp ^{i}\pt ^{\mu}\vp ^{j}}
\newcommand{\tetr}{\pt _{\mu}\vp ^{i}\pt ^{\mu}\vp ^{j}
                   \pt _{\nu}\vp ^{p}\pt ^{\nu}\vp ^{q}}   

\newcommand{\tden}{[(k+l+p)^{2}+\mu ^{2}](k^{2}+\mu ^{2})(l^{2}+\mu ^{2})}  

\begin{document}
\baselineskip=22pt
\begin{center}
\vspace{1.0cm}
{\large{\bf Covariant background field calculation of ultraviolet 
counterterms for the nonlinear sigma model in three dimensions$^{\ast}$}}
\end{center}
\vspace{1.5cm}
\baselineskip=12pt
\begin{center}
F.M. de Carvalho Filho$^{\dag}$ 
\end{center}
\begin{center}
{\it Center for Theoretical Physics\\
     Laboratory for Nuclear Science
     and Departament of Physics\\
     Massachusetts Institute of Technology\\
     Cambridge, Massachusetts\ \ 02139\ \ \ U.S.A.}
\end{center}
\baselineskip=15pt
\vspace{2cm}
\begin{center}
{\bf Abstract}
\end{center}
\vspace{0.2cm}

Based on the covariant background field method, we calculate the
ultraviolet counter\-terms up to two-loop order and discuss the
renormalizability of the three-dimensional non-linear sigma models
with arbitrary Riemannian manifolds as target spaces. We investigate
the bosonic model and its supersymmetric extension. We show that at
the one-loop level these models are renormalizable and even finite
when the manifolds are Ricci-flat. However, at the two-loop order, we
find non-renormalizable counterterms in all cases considered, so the
renormalizability and finiteness of such models are completely lost in
this order.

\vspace{\fill}

\noindent\makebox[66mm]{\hrulefill}

\footnotesize
$^{\ast}$This work is supported in part by funds provided by the U.S. 
Department of Energy(D.O.E) under cooperative agreement\#DE-FC02-94ER40818.


$^{\dag}$Email: farnezio@mitlns.mit.edu. 
Permanent address: Instituto de Ci\^encias, Escola Federal de 
         Engenharia de Itajub\'a, C.P. 50, Itajub\'a-M.G., Brazil. 
\normalsize
\newpage
\baselineskip=24pt
{\bf 1. \,Introduction}
\vspace{0.5cm}

Historically non-linear sigma models have been introduced by 
Schwinger\cite{1} and originally 
studied in the context of current algebra\cite{2}, but later, they have been 
proved an excelent theoretical laboratory, since in two-dimensional 
space-time they are similar to four-dimensional Yang--Mills theory\cite{3,4}. 
Afterwards, they have been used, in three dimensions, to study Fermi-Boson 
transmutation, anyonic 
high-$T_{c}$ superconductivity and related topics\cite{5}. These models have
also been helpful in the low-energy description of string(or superstring) 
theory\cite{6}. In this late case, they are defined in two dimensions and are
thus renormalizable. Indeed with the background field method\cite{7}
one is able to obtain the counterterms as functions of the gravitational 
fields\cite{8,9}. Moreover conformal invariance restraints these counterterms 
to zero, definning quantum corrections to Einstein equations\cite{6,10}. 
Hence, since in three dimensions they are models for 
membranes(or supermembranes)\cite{11}, in the same sense as the two-dimensional
cases are models for strings(or superstrings), it is natural and relevant 
to extend these methods to three space-time dimensions. The first step
in this direction is to investigate their renormalizability which 
as is well known may be achieved in the large $n$
expansion for $O(n)$ and $SU(n)$ invariant models\cite{12}. However,
perturbative calculations are essential in order that one may 
apply the above procedure.

In the present work we deal with the question of whether non-linear 
sigma models in
three-dimensional space-time may be perturbatively renormalizable or not.
As a matter of fact, 
such models are nonrenormalizable in the usual Dyson power-counting sense,
nevertheless it still may be possible to restore their renormalizability
cancelling all divergences by including higher 
terms in the Lagrangian and finding a counterterm to absorb every infinity.
So the main purpose of this paper is to calculate explicitly such counterterms 
and know what are and under which 
conditions generalized non-linear sigma models may be perturbative
renormalizable and even finite.
We carry out this program following the work of Alvarez-Gaum\'{e}, Freedman
and Mukhi\cite{9} and making calculations for the bosonic and supersymmetric
models on arbitrary manifolds, up to two-loop level. However, in our 
case, due to odd-dimensional nature of the space-time, 
the ultraviolet structure is quite different and special cares are 
taken throughout the calculations. We also examine some
particular cases such as those models defined on Ricci-flat and symmetric
spaces. We hope that the results obtained here can
be useful to implement the above mentioned method of extracting low-energy 
information in the membrane theory context, at least in a given order.

The paper is organized as follows. Section 2 contains the useful covariant 
background field for the bosonic three-dimensional
non-linear sigma model as well as the explicit calculations of the 
counterterms at the one- and two-loop level. In section 3 we present the
calculations throught two-loop order for the supersymmetric extension of
the previous model. In section 4 we discuss the results obtained and draw
our conclusions.
\vspace{1cm}

\noindent
{\bf 2. \,Calculation of the counterterms for the bosonic non-linear 
sigma model}
\vspace{0.5cm}

We shall begin studing the case of the three-dimensional purely bosonic 
non-linear sigma model defined by the action
\be
\act[\phi (x)]  = \frac{1}{2}\inte g_{ij}(\phi)\pt _{\mu}\phi^{i} \pt ^{\mu}\phi^{j}
\label{acao}
\ee
where the field $ \phi $ is a map from a three-dimensional Minkowski
space-time $ {\cal X} $(with
metric tensor $\eta ^{\mu \nu}$ such that $\eta ^{00}=-\eta ^{11}=
-\eta ^{22}=1$) taken as a base space to an arbitrary Riemannian 
manifold $ {\cal M} $ taken as target space and $ g_{ij}(\phi(x)) $ are the
components of the given metric tensor on $ {\cal M} $ .

As already mentioned in the Introduction, in the usual power-counting sense, 
the theory 
defined in equation (\ref{acao}) is not renormalizable. Therefore, we shall 
investigate the ultraviolet divergences and discuss the perturbative 
renormalizability of this model considering terms of higher dimensions, which
are discarded in the two-dimensional case\cite{9}, calculating explicitly 
all the counterterms needed.
In order to carry out this program we shall use here the
background field method\cite{7} explicitly covariantized via the normal
coordinate expansion\cite{13}. In fact, this procedure, which has been a powerful 
computational tool in Quantum Field Theory, allows us to computate
radiative corrections and the effective action in a manifestly covariant
way preserving the symmetries of the model under consideration\cite{14}.
For the non-linear sigma models it is already known\cite{8,9,15} and consists in
splitting the field $ \phi ^{i} $ into a classical(background) field 
$ \vp ^{i} $ and a quantum field $ \pi ^{i} $, taking, in followed, 
$ \pi ^{i} $ as a function of a new covariant quantum field $ \xi ^{i} $
in terms of which the normal coordinate is defined.  Moreover, using
the definitions, \,\,\,\, 
$ \xi ^{i} = e^{i}_{\,a}\xi^{a} , \,\, e^{i}_{\,a}e^{j}_{\,a} = g^{ij} , 
\,\,\, D^{\mu}\xi ^{a} = \pt ^{\mu}\xi ^{a} + \omm \pt ^{\mu}\vp ^{i}\xi^{b}
$,\,\,\,\,
\noindent
where $ e^{i}_{\,a} $ \,\,is a vielbein, \,\,$\omm $ is the spin 
connection of the manifold(with Latin indices $i,j,k,\cdots$) given by
\bear                                                                 
D_{i}e^{a}_{\,j}& \equiv &e^{a}_{\,j;i} = \pt _{i}e^{a}_{\,j} + \omm (e)e_{bj} -
\Gamma^{k}_{ji}e^{a}_{\,k} = 0 \nonumber\\
\omm &=& -e^{bj}\nabla _{i}e^{a}_{\,j} = -e^{bj}\pt _{i}e^{a}_{\,j} +
e^{bj}\Gamma^{k}_{ij}e^{a}_{\,k}
\ear    
and $\Gamma^{k}_{ij}$ the usual Christoffel symbol, one can move to tangent 
space(with Latin indices $a,b,c,\cdots ,h$)and get for the action (\ref{acao}) the following useable standard 
expansion(we refer to \cite{9} for details)
\be
\act[\vp + \pi ]=\act ^{(0)} [\vp ]+\act ^{(2)} [\vp ]+\act ^{(3)} [\vp ]+
\act ^{(4)} [\vp]+\cdots 
\nonumber 
\ee
\bear
&&\act ^{(2)}=\frac{1}{2}\inte \left\{ \left[ R_{iabj}(\vp )\doub -
A_{\mu}^{\,\,ac}A^{\mu cb}\right]\xi ^{a}\xi ^{b} + \pt_{\mu}\xi ^{a}
\pt ^{\mu}\xi ^{a} - \pve \xi ^{a}\stackrel{\leftrightarrow}{\pt _{\mu}}\xi ^{b} \right\}
\nonumber \\
&&\act ^{(3)}=\frac{1}{2}\inte \left\{ \left[ \frac{1}{3}D_{a}R_{ibcj}\doub + 
\frac{4}{3}R_{iabd}A_{\mu}^{\,\,dc}\pt ^{\mu}\vp ^{i} \right] \xi ^{a}\xi ^{b}
\xi ^{c} + \frac{4}{3}R_{iabc}\pt _{\mu}\vp ^{i}\xi ^{a}\xi ^{b}\pt ^{\mu}
\xi ^{c}\right\} \nonumber\\
&&\act ^{(4)}=\frac{1}{2}\inte \left\{ \left[ \frac{1}{3}D_{a}R_{ibce}A_{\mu}^{\,\,ed}
\pt ^{\mu}\vp ^{i} + \frac{1}{3}R_{eabf}A_{\mu}^{\,\,ec}
A^{\mu fd}+\frac{1}{12}D_{a}D_{b}R_{icdj}\doub\right.\right. + \nonumber\\
&&+\left.\frac{1}{3}R^{m}_{\,\,abi}R_{mcdj}
\doub \right] \xi ^{a} \xi ^{b} \xi ^{c} \xi ^{d} + \frac{1}{2}D_{a}R_{ibcd}\pt _{\mu}\vp ^{i}\xi ^{a} \xi ^{b} \xi ^{c}
\pt ^{\mu} \xi ^{d}+\nonumber\\
&&+\left. \frac{1}{3}R_{cabe}A_{\mu}^{\,\,ed} \xi ^{a} \xi ^{b}
\pt ^{\mu} \xi ^{c} \xi ^{d}+R_{cabd}A_{\mu}^{\,\,ec} \xi ^{a} \xi ^{b} \xi ^{c} \pt ^{\mu} \xi ^{d} + 
\frac{1}{3}R_{cabd} \xi ^{a} \xi ^{b} \pt _{\mu} \xi ^{c} 
\pt ^{\mu} \xi ^{d} \right\}
\label{expa}
\ear
where $R$ is the curvature tensor and $\pve \equiv \omm \pt _{\mu}\vp ^{i}$ 
a vector potential $\in SO(n)$. We have also omitted in the expression 
(\ref{expa}),the linear term in $\xi ,\, \act ^{(1)},$
since it vanishes as we use the equation of motion of $\vp $. Futhermore,
we are not interested here in renormalization of wave function for which it
could contribute. For computation up to two-loop order, the above results 
are all we need.

We are now able to compute the Feynman propagator for the $\xi$ field, which
is
\be
\langle 0\mid T\xi ^{a}(x)\xi ^{b}(y)\mid 0\rangle = \delta ^{ab}\Delta(x-y)
\label{prop}
\ee
\noindent
Before we start the calculation of the counterterms, a note is needed. We do
not use dimensional regularization in this work, since in odd dimensional
space-time it is in fact a renormalization prescription, which deletes all
divergent contributions automatically, rendering the theory finite. Thus, 
if we wish
to study the regularization effects in detail, we must make the
counterterm structure explicit. We therefore choose a Pauli--Villars
regularization (subtracting the infinites with the use of a regulator mass).
Using this gauge invariant procedure, the second and third diagrams of Fig. 1
(solid and double lines denote the $\xi$ and the background fields
respectively) do not contribute and the divergent one-loop counterterm 
arises from the first one, which is given by 

\be
D^{(1)} = R_{iabj}\doub \delta^{ab}\int \frac{d^{3}k}{(2\pi)^{3}}
\frac{i}{(k^2 - \mu^{2})} = \frac{\lb}{4\pi ^2} C_{aa}
\label{3}
\ee

\noindent
where $ C_{aa}\equiv R_{ij}\doub  , R_{ij}$ is the Ricci tensor and $\Lambda$
a cutoff introduced after a Wick rotation. We have also included in the 
propagator (\ref{prop})
a mass $\mu$ to avoid infrared divergences, which are not the issue of this
work.  

According to Friedan's interpretation\cite{8}, this counterterm may be absorved
in a redefinition of the metric as
\be
g^{ren}_{ij}=g_{ij} - \frac{\lb}{4\pi^2}R_{ij} \, .
\label{4}
\ee     
Thus we get an one-loop renormalized effective action

\be
\act _{eff}^{(1)}=\frac{1}{2}\inte \left(g_{ij}-\frac{\lb}{4\pi^2}C_{aa}\right).
\label{5}
\ee
\noindent
Futhermore, we can define a $\beta $ function as

\be
\beta _{ij}\equiv \frac{\delta g^{ren}_{ij}}{ \delta \lb}= - \frac{1}{4\pi^2}
R_{ij} .
\label{6}
\ee 
So we find that Ricci-flat manifolds are one-loop finite.

Now we shall perform the calculation of the two-loop counterterms. 
We have in this case several contributions and the diagrams are displayed 
in Fig. 2. However, we would like to appoint out that several 
vertices in the expansion(\ref{expa}) contain derivatives and as a consequence there 
will be a momentum flow through each vertice becoming such diagrams much 
more complicated, so an special attention
will be paid in this calculation. 
Moreover, in order to define a renormalization procedure, we shall use here the
BPHZ-type scheme making Taylor subtractions around zero external
momenta\cite{16}. The integrals appearing in this approach will evaluated, 
when necessary, by applying the Feynman parametrization method and using the 
Ref. \cite{17}.

For the first diagram, we have the square of an one-loop diagram, which is
easily computable and the result is
\bear
D^{(2a)}&=&\frac{\lb^2}{4\pi^4}\left[\frac{1}{4}[D_{a}R_{ib}-3D_{b}R_{ia}]
\om _{j}^{\,ab}+\frac{1}{6}R_{ac}\omm \om _{j}^{\,bc}+\frac{1}{3}R_{c(ab)d}
\om _{i}^{\,cb}\om _{j}^{\,da}+\frac{1}{12}D^{a}D_{i}R_{aj}-\right. \nonumber\\
&-&\left.\frac{1}{8}D_{a}D^{a}R_{ij}+\frac{1}{4}R_{iabc}R_{j}^{\,\,abc}+
\frac{1}{6}R_{ia}R_{j}^{\,a}\right]\doub-\frac{\lb ^4}{72\pi ^{4}}R \, ,
\label{cosm}
\ear
where the symbol $(\,)$ in tensor indices denotes symmetrization. Also,
in the above expression, the Bianchi and cyclic indenties have been used.
The contribution (\ref{cosm}) implies still a redefinition of the vacuum. It is
in essence a one-loop counterterm.
Actually, althoug not in the usual sense, since the $g_{ij}$ does not couple to
a spatial derivative, it corresponds to a {\it cosmological term}.

The above procedure applied for diagrams (2b) and (2c) gives
\bear
D^{(2b+2c)}&=&\frac{1}{192\pi^{4}}\left(\frac{\pi}{3}\frac{\lb ^{3}}{\mu}+
4\lb ^{2}\right)
(R_{ab}\om _{i}^{\,ca}\om _{j}^{\,cb}+R^{ab}R_{iabj}) \doub + \nonumber\\
&+&\frac{1}{32\pi ^{3}}\frac{\lb}{\mu}X_{ijpq}\tetr ,
\label{8}
\ear
where
\bear
X_{ijpq}&\equiv &\frac{1}{4}\left[(D_{a}R_{ic}+D_{i}R_{ac}-2D_{c}R_{ia})
R_{p(cb)q}+(D_{b}R_{idca}+D_{b}R_{idca}+\right. \nonumber\\
&+&\left. D_{d}R_{ibca})R_{p(dc)q}\right]\om_{j}^{\,ab}+
\frac{1}{3}R_{a(bc)d}R_{p(be)q}(\om _{i}^{\,ac}\om _{j}^{\,de}+
\om _{i}^{\,ae}\om _{j}^{\,dc})-\frac{1}{2}R_{ab}R_{pcdq}
\om _{i}^{\,ac}\om _{j}^{\,bd}  \nonumber\\
&+&\frac{1}{2}R_{abcd}R_{pbcq}\om _{i}^{\,ae}\om _{j}^{\,de}+ 
\frac{1}{48}\left[2D^{2}R_{iabj}R_{pabq}+(4D_{a}D_{i}R_{jb}-5D_{a}D_{b}
R_{ij})R_{pabq}\right]+ \nonumber\\
&+&\frac{1}{3}\left[R_{abci}\left(R_{a(bd)j}R_{p(dc)q}+R_{a(ec)j}R_{p(be)q}
\right)-R_{ai}R_{abcj}R_{p(bc)q}\right]\, \, . \nonumber
\ear

\noindent
Next we are going to consider the contribution (2d) which is really new
since it contains the first {\it non-renormalizable} counterterm. We
divide it into two pieces(2d/1 and 2d/2). The first, analogous to the previous, 
is 
\be
D^{(2d/1)}=\frac{1}{2(2\pi)^6}Y_{ijpq}\tetr \int \dint\frac{1}{\tden}
\label{9}
\ee
where, 
\bear
Y_{ijpq}&\equiv & \frac{1}{18}\left[(D_{a}R_{pbcq}+D_{b}R_{pacq}+
D_{c}R_{pabq})(D_{a}R_{ibcj}+4R_{iabd}\om_{j}^{\,dc})+4(2D_{a}R_{p(bc)q}+
\right. \nonumber\\
&+&\left. D_{c}R_{pabq})R_{i(ab)d}\om_{j}^{\,dc}+
16R_{iabd}(R_{p(ab)e}\om_{j}^{\,dc}\om_{q}^{\,ec}+R_{p(ac)e}\om_{j}^{\,dc}
\om_{q}^{\,eb}+R_{p(bc)i}\om_{j}^{\,dc}\om_{q}^{\,ea})\right]\,.\nonumber\\
\ear

The second one, also non-renormalizable, is given by the following 
expression in momentum space:
\be
D^{(2d/2)}=\frac{2}{9(2\pi)^6}R_{iabc}(R_{j}^{\,\,abc}+
R_{j}^{\,\,bac})\pt ^{\mu}\vp ^{i}
\pt ^{\nu}\vp ^{j}\inta \frac{[2l_{\mu}l_{\nu}+l_{\mu}(k+p)_{\nu}]}{\tden}
\label{10}
\ee
leading to higher derivative counterterms, that is,
\bear
D^{(2d)}&=&-\frac{1}{192\pi ^4}\left\{ \left[8\lb ^2-\frac{1}{5}\pi ^2 
\left(ln\frac{\lb ^2}{\mu ^2}-\frac{8}{3\pi}\frac{\lb}{\mu}-
\frac{4}{3\pi ^2}\frac{\lb ^2}{\mu ^2}\right)p^2\right]\eta ^{\mu \nu}+ \right.
\nonumber\\
&+&\left. \frac{1}{5}\pi ^{2}\left(3ln\frac{\lb ^2}{\mu ^2}-
\frac{4}{3\pi}\frac{\lb}{\mu}-\frac{4}{3\pi ^2}\frac{\lb ^2}{\mu ^2}\right)
p^{\mu}p^{\nu}\right\}R_{iabc}R_{j}^{\,\,abc}\pt _{\mu}\vp ^{i}\pt _{\nu}\vp ^{j}-\nonumber\\
&-&\frac{1}{64\pi ^2}ln\frac{\lb ^2}{\mu ^2}Y_{ijpq}\tetr \, .
\label{11}
\ear

We have computed explicitly the counterterms in the following cases: \\
\noindent
{\it (i)  Ricci-flat spaces} ($R_{ij}=0$); \\
\noindent
{\it (ii) Locally symmetric spaces}, where $D_{i}R_{jklm}=0$ \, , which
include the the $O(n)$ and $CP^{n-1}$  models.

In the first case we have finiteness at one loop level, but nonrenormalizability
at two loops; the counterterm, which is given by
\bear
D^{(2)}_{R-f}&=&D^{(2d)}-\frac{\lb ^2}{8\pi ^4}\left(R_{dabc}
\om_{i}^{\,db}\om_{j}^{\,ac}-\frac{1}{2}R_{iabc}R_{j}^{\,\,abc}\right)\doub +
\nonumber\\
&+&\frac{1}{32\pi ^3}\frac{\lb}{\mu}X_{ijpq}(R_{ab}=0)\tetr
\label{12}
\ear
\noindent
is not of the form of the original Lagrangian. 

In the second case we have a
renormalizability at the one-loop order, but some infinites still remain at the 
two-loop order. Specifically, in the $O(n)$ non-linear sigma model, where the 
metric, curvature and Ricci tensors are, respectively, given by
\bear
g_{ij}(\vp) = \delta _{ij}+\frac{\vp _{i}\vp _{j}}{1-|\vp|^2} \, , \,\,\,\,
R_{ijkl} = g_{ik}(\vp)g_{jl}(\vp)-g_{il}(\vp)g_{jk}(\vp) \, , \,\,\,\,
R_{ij} = (n-2)g_{ij}
\ear
we have, at the one-loop level,
\be
\delta \lag ^{(1)}_{O(n)}=\frac{1}{4\pi ^2}(n-2)\lb g_{ij}(\vp)\doub \, ,
\label{oene}
\ee       
and, at the two-loop order,
\newpage

\bear
\delta\lag ^{(2)}_{O(n)}&=&-\frac{1}{48\pi ^4}\left\{\left[(3n-n^2+4)
+\frac{\pi}{12}(n^2-5n+6)\frac{\lb ^3}{\mu}\right]\eta ^{\mu
\nu}+\frac{\pi ^2}{10}(n-2)\times\right.\nonumber\\
&\times&\left.\left[(3ln\frac{\lb ^2}{\mu ^2}-\frac{4}{3\pi}
\frac{\lb}{\mu})p^{\mu}p^{\nu}- (ln\frac{\lb ^2}{\mu ^2}-
\frac{8}{3\pi}\frac{\lb}{\mu}-\frac{4}{3\pi ^2}\frac{\lb ^2}{\mu ^2})
p^2 \eta ^{\mu \nu}\right]\right\}g_{ij}\pt _{\mu}\vp ^{i}\pt _{\nu}\vp ^{j} +
\nonumber\\ 
&+&\left\{\frac{1}{96\pi^3}\frac{\lb}{\mu}\left[(n^2-3n-3)g_{ij}g_{pq}+
(n+3)g_{ip}g_{jq}\right]-\frac{(n+1)}{72\pi ^2}ln\frac{\lb ^2}{\mu^2}
g_{ij}g_{pq}\right\}\times\nonumber\\
&\times&\tetr \,\, .  
\label{oened}
\ear

Therefore, we can easily see from (\ref{oene}) and (\ref{oened}) that the $O(n)$ model is 
renormalizable at one-loop order and obviously finite whether $n=2$. 
Nevertheless, it remains 
non-renormalizable at the two-loop level since the quartic term in the 
field $\vp$ in (\ref{oened}) is not
completely cancelled, even whether we consider the $O(2)$ case where some 
important cancellations are obtained.

In the next section we shall go on to consider the case of the fermions and
write down the corresponding countertems again up to two-loop order.
\vspace{1.0cm}

\noindent
{\bf 3. \,Supersymmetric extension}
\vspace{0.5cm}

We shall consider now the supersymmetric extension of the action (\ref{acao}), 
which reads
\be
S=\frac{1}{4i}\inte d^2 \theta g_{ij}(\Phi ^{k})\overline{D\Phi} ^{i}
D\Phi ^{j} ,
\label{aca}
\ee
\noindent
where  $D\Phi^{i}$ is the supercovariant derivative of $\Phi^{i}$ , 
such that,
$D_{\alpha}=\frac{\pt}{\pt\bar{\theta}^{\alpha}}-i(\partiall
\theta)_{\alpha}$  ; being $\theta$ a two-component Majorana anticommuting
variable and $\Phi ^{i}$ a scalar superfield whose expansion in terms of the
component fields $\phi ^{i}$ (the scalar fields), $\psi^{i}$ (the Majorana 
spinor which are the fermionic partners) and $F^{i}$ (the auxiliary fields), 
in the Majorana represention for the gamma matrices, is given by
\be
\Phi ^{i}=\phi ^{i}(x)+\bar{\theta}\psi ^{i}(x)+\frac{1}{2}\bar{\theta}\theta
F^{i}(x).
\label{sup}
\ee

Substituting the above relations back into  (\ref{aca})  , integrating over the Grassmann
variable $\theta$ by means of the standard rules of Berezin integration and
eliminating the auxiliary fields, we finally obtain the Lagrangian
in terms of component fields
\be
L=\frac{1}{2}\left[g_{ij}(\phi)\pt _{\mu}\phi ^{i}\pt ^{\mu}\phi ^{j}+
ig_{ij}(\phi)\psb ^{i}\gamma ^{\mu}D_{\mu}\psi ^{j}+\frac{1}{6}R_{ijkl}
(\psb ^{i}\psi ^{k})(\psb ^{j}\psi ^{l})\right] \,\, ,
\ee
where
\be
(D_{\mu}\psi)^{j}=\pt _{\mu}\psi ^{j}+\Gamma _{kl}^{j}\pt _{\mu}\phi ^{k}
\psi ^{l} \,\, .
\ee
The background field method works well as in the previous case\cite{9}. We consider
the Fermi fields $\psi ^{i}$ to be quantum fields, avoiding background quantum
splitting for anticommuting variables. We obtain
\be
g_{ij}(\phi)\psb ^{i}\DI\psi ^{j}=\left(g_{ij}(\vp)+\frac{1}{3}R_{iklj}\right)\xi ^{k}
\xi ^{l}\psb ^{i}\DI\psi ^{j}+\frac{1}{2}R_{ijkl}\pt _{\mu}\vp ^{l}
\xi ^{k}(\psb ^{i}\gamma ^{\mu}\psi ^{j}) \,\, .
\ee

We can now write all relevant objects in terms of tangent space variables,
using
\be
\xi ^{a}=e_{i}^{\,a}\xi ^{i} \, , \,\,\, \psi ^{a}=e_{i}^{\,a}\psi ^{i} \, ,
\,\,\, (D_{\mu}\psi)^{a}=\pt _{\mu}\psi ^{a}+\omm \pt _{\mu}\phi ^{i}
\psi ^{b} \,\, .
\ee
Gathering together all relevant informations, we obtain
\be
S=S^{(0)}[\psi]+S^{(1)}[\psi]+S^{(2)}[\psi]+S^{(3)}[\psi]+S^{(4)}[\psi]+\cdots
\ee
being
\bear
S^{(0)}&=&\frac{1}{2}\inte \psb ^{a}(i\gamma ^{\mu}\pt _{\mu}-\mu)\psi ^{a} \,\,
,\,\,\,\, \,\,\,\,\,\,\,\,\,S^{(1)}=\frac{1}{2}\inte i\psb ^{a}\gamma ^{\mu}\pve \psi ^{b} \,\, ,
\nonumber\\
S^{(2)}&=&\frac{1}{6}\inte R_{acdb}\xi ^{c}\xi ^{d}i\psb ^{a}\gamma ^{\mu}
D_{\mu}\psi ^{b} \,\, , \,\,\,
S^{(3)}=\frac{1}{4}\inte R_{abci}\pt _{\mu}\vp ^{i}\xi ^{c}i\psb ^{a}
\gamma ^{\mu}\psi \,\, , \nonumber\\
S^{(4)}&=&\frac{1}{12}\inte R_{abcd}\psb ^{a}
\psi ^{c}\psb ^{b}\psi ^{d} \,\, ,
\ear
where a mass has been introduced again in order that we obtain infrared
finite results. Using the Pauli--Villars regulator, we get a vanishing result
at one-loop(see Fig. 3, with the dashed lines denoting the
fermion propagators). In this calculation, the infrared regulator and the 
ultraviolet cutoff
were taken to be equal to those ones of the bosonic case($\mu$ and $\lb $
respectively) in order not to introduce an explicit breaking of supersymmetry.
In the following we shall use this same procedure. Now, at the two-loop order, 
we have the contributions 
shown in Fig. 4. In the first diagram we have a contribution arising from
$S^{(2)}[\psi]$  , given by
\be
\delta S^{(4a)}[\psi]=-\frac{1}{6}\inte R_{acdb}\langle T[\xi ^{c}
\xi ^{d}\psb ^{a}\gamma ^{\mu}(\pt _{\mu}\psi ^{b}+iA_{\mu}^{\,\,bc}\psi ^{c})]
\rangle
\ee
Upon contracting the $\xi$ and the $\psi$ fields, we obtain
\be
\delta S^{(4a)}=\frac{\lb ^4}{36\pi ^4}\inte R(x)
\ee
which is analogous to previous computations(see Eq. (\ref{cosm})). Nevertheless, 
due to a factor of $2$, there is no cancellation between these terms. Note that
diagrams (b) and (c) do not contribute. Finally, the last contribution is

\be
\delta S^{(4d)}=-\frac{1}{32}\inte d^3 yR_{abcd}(x)\pt _{\mu}\vp ^{d}(x)
R_{efgh}(y)\pt _{\nu}\vp _{\nu}\vp ^{h}(y)\langle T[(\xi ^{c}\psb ^{a}\gamma ^{\mu}\psi ^{b})(x)(\xi ^{g}\psb ^{e}
\gamma ^{\nu}\psi ^{f})(y)]\rangle \nonumber
\ee
which, after a lengthy calculation, is given by

\bear
\delta S^{(4d)}&=&\frac{1}{192\pi ^4}\left\{ \left[ \lb ^2 -\frac{1}{5}
\pi ^2(3ln\frac{\lb ^2}{\mu ^2}-\frac{4}{\pi}\frac{\lb}{\mu}+\frac{8}{3\pi ^2}
\frac{\lb ^2}{\mu ^2})p^2\right]\eta ^{\mu\nu}\right. +\nonumber\\
&+&\left. \frac{2}{15}\pi ^2\left[ln\frac{\lb ^2}{\mu ^2}+
\frac{1}{4\pi ^2}\right]p^{\mu}p^{\nu}\right\}\inte R_{iabc}
R_{j}^{\,\,abc}\pt _{\mu}\vp ^{i}\pt _{\nu}\vp ^{j} \,\, .
\ear
 Therefore, even though we have a finiteness at the one-loop level
the supersymmetric extension is not sufficient to remove completely
all those non-renormalizable
counterterms at two-loop or higher order. This result is crucially different from
that one in two dimensions. In two space-time dimensions non-linear sigma 
models have no two- or three-loop terms in their $\beta$-function on 
any target manifold\cite{8,9}. 
\vspace{1.0cm}

\noindent
{\bf 4. \,Conclusions}
\vspace{0.5cm}

In this paper we have calculated the divergences in the  effective action
of the three-dimensional non-linear sigma models and determined their
one- and two-loop counterterms by the covariantized background field method.
We have shown that at the one-loop order all the divergences may be absorbed
in renormalizable counterterms. In fact, in some particular cases, 
when we consider the
supersymmetric extension or symmetric and Ricci-flat manifolds, 
such models are even finite. On the other hand, at the two-loop level, we
have found non-renormalizable counterterms in all cases considered so that 
the one-loop renormalizability and finiteness are completely lost. For
instance, in the supersymmetric case, cancellation between bosons and 
fermions is not enough to render the model renormalizable . We think that 
the same continues
to be true for higher supersymmetry(we have been working explicitly the 
case $N=2$). Restrictions of the manifold may result in the fact some 
counterterms might be zero, but not all of them.
However, we believe that these  two-loop results, though negative in the 
sense that we do not find any sensible renormalizable theory in any simple
case, should not be used to discard the models studied so far. Actually, 
it is possible to extract physical information out of the results obtained
at the one-loop level which can be important in view of the many applications 
of sigma models, as we have already mentioned in the Introduction of this 
paper. By the way, the quantum gravity is a well known
case of a non-renormalizable theory whose divergences, at the one-loop order, 
can be absorbed in the counterterms leading a meaningful theory\cite{18}.

Moreover, it seems also important for us to point out the different result 
one obtains from the
perturbation theory used here, and other results based on the
large $n$ behaviour\cite{12,19}, which define a renormalizable theory. 
Indeed, from $1/n$ perturbation of the $CP^{n-1}$ 
model one learns that the model
has two phases, one having a massive 
$n$-plet and a massless abelian gauge field, and another with a massless 
$(n-1)$-plet and a gauge field displaying no pole in the propagator.
In these sigma models, cancellation of divergences are a consequence of the
definition of the auxiliary field propagator\cite{20}, and the identity shown in
Fig. 5.

We should also make some remarks concerning general four-dimensional non-linear
sigma models. Although already studied many years ago\cite{15}, it is not difficult
to obtain the first few counterterms using the background field method. Indeed,
the Lagrangian
\be
L=g_{ij}\psb ^{i}\DI\psi ^{j} + g_{ij}D_{\mu}\vp ^{i}D^{\mu}\vp ^{j}
\nonumber
\ee
has a background-quantum expansion given by
\be
L=L_{cl}(\vp ^{a},\psi ^{a}) + R_{iabj}\left(\doub + \frac{1}{3}\psb ^{i}\DI\psi
^{j}\right)\xi ^{a}\xi ^{b} \nonumber
\ee
with a gauge field $\pve = \omm \pt _{\mu}\vp ^{i}$ \, .
The diagram with two, three and four $\pve $ legs cannot be made to vanish,
and we need a counterterm $F_{\mu \nu}^{2}$ \, , which is non-renormalizable
already at the one-loop level. 

Therefore, we are led to conjecture that, in all case considered, several 
of the infinites we found are fake infinites
produced by perturbation theory, or else. As matter of fact, we believe
that the theory may also have different phases(not those ones above mentioned) 
and that such infinites, 
as already noted by several authors in other contexts\cite{21}, may have
nothing to do with the physical content of the models investigated. This calls
for an explanation.

Finally, we would like to mention that more recently considerable 
discussion about non-renormalizable interactions has been done\cite{22,23,24}
and even certain approaches
for the corresponding theories have been proposed\cite{23,24}. In particular, 
J. Gegelia et al.\cite{24} have developed a method to extract physical information out
of the series of non-renormalizable theories which coincides with the usual
renormalization procedure(in terms of counterterms) for renormalizable ones.
We hope that our calculations as well as these recent works can be useful
for our understanding of this subject.                                                  
\newpage

\noindent
{\bf Acknowledgements}

I wish to thank Professor Elcio Abdalla for collaboration in early
stages of this work, helpful discussions and reading of 
the manuscript. 
I would also like to thank the Center for Theoretical 
Physics at MIT for hospitality. This work was partially supported by
CAPES(Brazil).
\medskip


\newpage

\bigskip
\noindent
{\bf Figure Captions}

\bigskip
\bigskip
\bigskip
\normalsize
\noindent
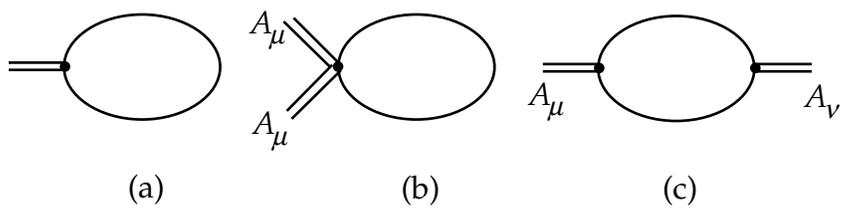
Figure 1. One-loop order contributions.

\bigskip
\bigskip
\noindent
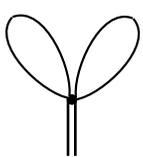
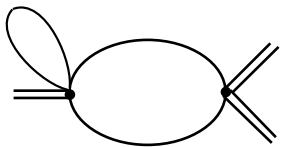
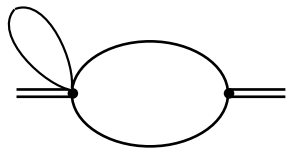
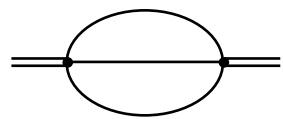
Figure 2. Two-loop contributions.

\bigskip
\bigskip
\noindent
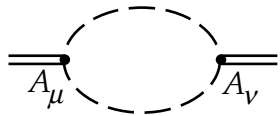
Figure 3. Vanishing contribution upon use of gauge invariant regularization.

\bigskip
\bigskip
\noindent
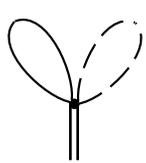
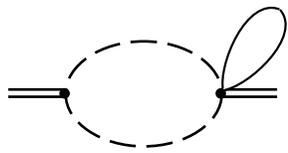
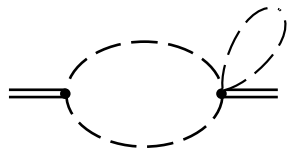
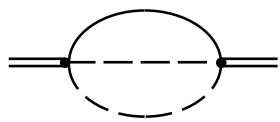
Figure 4. Two-loop contribution for the supersymmetric case.

\bigskip
\bigskip
\noindent
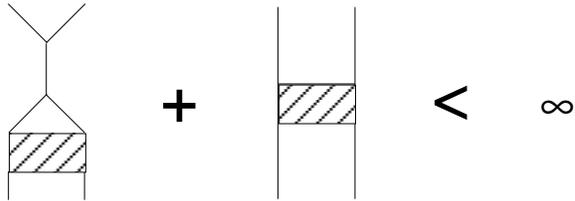
Figure 5. Cancellation mechanism in the $1/n$ expansion.
\end{document}